# Lorentz transformation directly from the invariance of the speed of light via the addition law of parallel speeds

Bernhard Rothenstein, "Politehnica" University of Timisoara, Department of physics, Timisoara, Romania

*Abstract. We show that starting with the addition law of parallel speeds derived as a consequence of the invariance of the speed of light, the Lorentz transformations for the space-time coordinates can be derived.*

## 1. The case of the addition law of parallel speeds derived directly from the invariance of c

Mermin[1] and Peres[2] have given a derivation of the relativistic addition law of parallel speeds, which is an immediate consequence of the invariance of the speed of light c in empty space. The relativistic arsenal (Lorentz transformation (LT), time dilation, length contraction or the relativity of simultaneity is not used. Many other derivations of the relativistic addition law of parallel speeds, without using the LT, based on formulas which account for time dilation and length contraction are known.[3]

We have presented[4] some time ago a derivation of the LT based on the addition law of parallel speeds and so from the invariance of c, considering its derivation in[1,2]. Kapuscik[5] considered that the derivation we proposed should take into account the time dilation effect formula. We revisit the derivation we have proposed, considering that it has some pedagogical potential, showing that we can avoid the suggestion made in[5].

Consider a bullet that moves with speed u relative to an inertial reference frame **I** and with speed u' relative to an inertial reference frame **I'**. The two inertial reference frames are in the standard arrangement, **I'** moving with constant speed V relative to I all showing in the positive direction of the overlapped OX(O'X') axes. The three mentioned speeds are related by[1,2]

$$u = \frac{u' + V}{1 + \frac{Vu'}{c^2}} \qquad (1)$$

and by

$$u' = \frac{u - V}{1 - \frac{Vu}{c^2}}. \qquad (2)$$

By definition

$$u = \frac{dx}{dt} \qquad (3)$$

and

$$u' = \frac{dx'}{dt'}. \tag{4}$$

During its motion, the bullet generates the event E(x,t) when detected from **I** and E'(x',t') when detected from **I'**. With (3) and (4), (1) can be presented as

$$\frac{dx}{dt} = \frac{dx' + Vdt'}{dt' + \frac{V}{c^2}dx}. \tag{5}$$

For the same reasons (2) can be presented as

$$\frac{dx'}{dt'} = \frac{dx - Vdt}{dt - \frac{V}{c^2}dx}. \tag{6}$$

Eq.(5) suggests that

$$dx = f_1(V)(dx' + Vdt') = f_1(V)dx'(1 + \frac{V}{u'}) \tag{7}$$

$$dt = f_1(V)(dt' + \frac{V}{c^2}dx') = f_1(V)dt'(1 + \frac{Vu'}{c^2}) \tag{8}$$

$$dx' = f_2(V)(dx - Vdt) = f_2(V)dx(1 - \frac{V}{u_x}) \tag{9}$$

$$dt' = f_2(V)(dt - \frac{V}{c^2}dx) = f_2(V)dt(1 - \frac{Vu}{c^2}) \tag{10}$$

where $f_1(V)$ and $f_2(V)$ are functions of the relative speed V but not of the coordinates of the events involved in the transformation process, in accordance with the hypothesis of linearity.

As in all the strategies adopted to derive the LT we should prove that the equations we have derived so far keep the speed of light invariant. Consider a source of light S(0,0) located at the origin O of **I**. At the origin of time in **I** and in **I'** (t=t'=0), when the origins O and O' are located at the same point in space, S(0,0) emits a light signal in the positive direction of the overlapped axes. The observers of the two frames measure the speed of that light signal. If the speed of light c is an invariant then we should have

$$\frac{dx}{dt} = c \tag{11}$$

and

$$\frac{dx'}{dt'} = c. \tag{12}$$

The same light signal synchronizes the clocks of **I** (C(x,0)) and of **I'** (C'(x'0)) in accordance with a clock synchronization procedure proposed by Einstein[6]. The clocks $C_0(0,0)$ and $C'_0(0,0)$ located at the origins O and O' respectively are of particular interest. When detected from **I** clock $C'_0(0,0)$ measures a proper time interval $dt^0$, its displacement relative to **I'** is equal to zero (dx'=0), clocks of I measuring a non-proper time interval dt. In accordance with (2) we have

$$dt = f_1(V)dt^0. \tag{13}$$

Detected from **I'** clock $C_0(0,0)$ moves with speed –V in the negative direction of the overlapped axes and its displacement detected from **I** is equal to zero (dx=0) and measures a proper time interval $dt^0$. The synchronized clocks of I' measure a non-proper time interval dt' and (10) leads to

$$dt' = f_2(V)dt^0. \tag{14}$$

Symmetry requires dt=dt' because otherwise the experiment described above could enable to detect, confined in an inertial reference frame, if we are in a state of rest or in a state of uniform motion. The result is

$$f_1(V) = f_2(V) = f(V). \tag{15}$$

With all that (7),(8),(9) and (10) become

$$dx = f(V)dx'(1+\frac{V}{u'}) \tag{16}$$

$$dt = f(V)dt'(1+\frac{Vu'}{c^2}) \tag{17}$$

$$dx' = f(V)dx(1-\frac{V}{u}) \tag{18}$$

$$dt' = f(V)dt(1-\frac{Vx}{c^2}). \tag{19}$$

Multiplying (16) and (18) side by side we obtain

$$f^2(V)(1+\frac{V}{u'})(1-\frac{V}{u}) = 1 \tag{20}$$

whereas multiplying (17) and (19) side by side we obtain

$$f^2(V)(1+\frac{Vu'}{c^2})(1-\frac{Vu}{c^2}) = 1. \tag{21}$$

The first result is the relativistic identity

$$\left(1+\frac{V}{u'}\right)\left(1-\frac{V}{u}\right) = \left(1+\frac{Vu'}{c^2}\right)\left(1-\frac{Vu}{c^2}\right). \tag{22}$$

Expressing the right sides of (20) and (21) as a function of u or u' only we obtain

$$f(V) = \frac{1}{\sqrt{1-\frac{V^2}{c^2}}} \tag{23}$$

and so the LT we are looking for are

$$dx = \frac{dx' + Vdt'}{\sqrt{1-\frac{V^2}{c^2}}}. \tag{24}$$

$$dt = \frac{dt' + \frac{V}{c^2}dx'}{\sqrt{1-\frac{V^2}{c^2}}}. \tag{25}$$

We obtain in a similar way the inverse transformations.

Consider the events $E_0(0,0)$ and $E'_0(0,0)$ associated with the fact that the origins O and O' are located at the same point in space. The events generated by the bullet are E(x,t) when detected from **I** and E'(x',t') when detected from **I**. Being generated by the same moving bullet the two events take place at the same point in space when the clocks C(x,0) and C'(x,0) of the two frames located at that point read t and t' respectively E(x,t) and E'(x',t') represent by definition the same event. The distance between events E(x,t) and $E_0(0,0)$ is

$$\Delta x = x - 0 \tag{26}$$

being separated by a time interval

$$\Delta t = t - 0. \tag{27}$$

For the same reasons we have in **I'**

$$\Delta x' = x' - 0 \tag{28}$$
$$\Delta t' = t' - 0. \tag{29}$$

As a consequence we can express the transformation equations as a function of the space-time coordinates of the events involved in the transformation process

$$x = \frac{x' + Vt'}{\sqrt{1-\frac{V^2}{c^2}}}. \tag{30}$$

$$t = \frac{t' + \frac{V}{c^2}x'}{\sqrt{1-\frac{V^2}{c^2}}}. \tag{31}$$

Revisiting our first presentation of that approach[4], Kapuscik claimed that in order to find out the function f(V) we should know the formula which accounts for the time dilation effect $t = \dfrac{t'}{\sqrt{1-\dfrac{V^2}{c^2}}}$. As we have shown that is not the case.

**2. The case of the addition law of parallel speeds from the time dilation formula**

Mathews[3] and many others[6,7,8] derive the addition law of parallel velocities based on the time dilation and length contraction formulas. We show that it leads directly to the Lorentz transformation for the space-time coordinates of the same event.

Consider a third inertial reference frame **I⁰ and** a clock $C_0^0(0,0)$ located at its origin $O^0(0,0)$. It measures a proper time interval $\Delta t^0$. With adequate initial conditions it could also represent its reading $t^0$. Reference frame **I⁰** moves with constant speed u relative to **I**, moves with speed u' relative to **I'** and as we know **I'** moves with speed V relative to **I.** Observers from **I** consider that the change in the reading of the clock $C_0^0(0,0)$ is $\Delta t$ related to $\Delta t^0$ by the time dilation formula

$$\Delta t = \dfrac{\Delta t^0}{\sqrt{1-\dfrac{u^2}{c^2}}} \qquad (32)$$

For the same reasons, observers from **I'** consider that the change in the reading of the clock $C_0^0(0,0)$ is $\Delta t'$ related to $\Delta t^0$ by

$$\Delta t' = \dfrac{\Delta t^0}{\sqrt{1-\dfrac{u'^2}{c^2}}}. \qquad (33)$$

Expressing the right side of (32) as a function of u' via (1) and taking into account (33) we obtain

$$\Delta t = \dfrac{\Delta t^0}{\sqrt{1-\dfrac{V^2}{c^2}}\sqrt{1-\dfrac{u'^2}{c^2}}}\left(1+\dfrac{Vu'}{c^2}\right) = \dfrac{\Delta t' + V\Delta x'}{\sqrt{1-\dfrac{V^2}{c^2}}}. \qquad (34)$$

In accordance with (11) and (12) Eq. (34) leads to (24).

**3. Conclusions**

The Lorentz transformation could be derived from the addition law of relativistic speeds derived without using it and that the time dilation formula, obtained as a result of a thought experiment could lead to the Lorentz transformations using the addition law of speeds.